\documentclass[prd,%twocolumn,%
preprint,
superscriptaddress,nofootinbib,%
tightenlines ]{revtex4}
\usepackage{epsfig}
\usepackage{amssymb}
\usepackage{color}
\usepackage{hyperref}

\newcommand{\ben}{\begin{displaymath}}
\newcommand{\een}{\end{displaymath}}
\newcommand{\be}{\begin{equation}}
\newcommand{\ee}{\end{equation}}
\newcommand{\bea}{\begin{eqnarray}}
\newcommand{\eea}{\end{eqnarray}}
\begin{document}
%\preprint{MKPH-T-06-16}
\title{Critical comments on the quantization of the angular momentum:\\
II. Analysis based on the requirement that the eigenfunction of the third component of the operator of the angular momentum must be a single valued periodic function}
\author{G.~Japaridze}
\affiliation{Clark Atlanta University, Atlanta, GA, USA}
\affiliation{Kennesaw State University, Kennesaw, GA, USA}
\author{A.~Khelashvili}
\affiliation{Institute of High Energy Physics, Iv. Javakhishvili Tbilisi State University, Tbilisi, Georgia}
\author{K.~Turashvili}
\affiliation{Institute of High Energy Physics, Iv. Javakhishvili Tbilisi State University, Tbilisi, Georgia}

%\acute{3 October, 2019}

\begin{abstract}
\noindent We discuss the requirement of single valuedness and periodicity of eigenfunction of the third component of the operator of angular momentum. This condition, imposed on a non observable, is often used to derive that the eigenvalues of angular momentum could be only integer. We re-examine the arguments based on this requirement and alternate condition imposed by Pauli and show that they do not follow from the first principles and therefore these constraints can dropped. Consequently, we arrive to the same conclusion as in \cite{JKT}: there exist regular, normalizable eigenfunctions with the non-integer eigenvalues thus a non-integer angular momentum is perfectly admissible from the theoretical viewpoint.  The issue of the nature of eigenvalues forming the spectrum of the angular momentum remains open. 
What 
can be derived from the first principles is that to a fixed value of the angular momentum $L$  corresponds a discrete spectrum of eigenvalues of the third component of the angular momentum, $m$, defined by the relation $|m|=L-k,k=\{0,1,\cdots , [L] \}$, where $[L]$ is an integer part of $L$.\\
\noindent As a mathematical byproduct of our analysis of eigenfunctions, we present an alternate definition of a power of a complex number allowing to retain initial translational invariance of a base.
%This relation is the result of a physical requirement of normalizability of a wave functio
\end{abstract}


%\pacs{
%03.65.Nk,
%Nonrelativistic scattering theory
%11.10.Gh,
%Renormalization
%12.39.Fe.}
%Chiral Lagrangians
%04.60.Ds  Canonical quantization
%04.60.Gw  Covariant and sum-over-histories quantization
%03.70.+k  Theory of quantized fields
%          (see also 11.10 Field theory)

%\pacs{ ??xxx, xyy, xyz?? }

%\pacs{

\maketitle

\section{introduction}
As pointed out in our first publication \cite{JKT}, the issue whether the spectrum of the angular momentum consists of only integer numbers (in units of Planck's constant $\hbar$; throughout $\hbar=1$) is analyzed using three different methods, each of them based on specific requirement either on eigenfunctions or on algebra of commutation relations. All three approaches yield the same result: the eigenvalues of the angular momentum and its third component can be only integer \cite{shiff}-\cite{Weinberg2015}.

In \cite{JKT} we examined approach  based on two methods, namely:
\begin{itemize}
\item Integer spectrum of eigenvalues follows from the requirement that the eigenfunctions of the operator of angular momentum must be regular, i.e. normalizable 
\item Integer spectrum of eigenvalues follows from the commutation relations of the operators of physical quantities 
\end{itemize}
and showed that  neither of these methods guarantee that the spectrum consists of only integers.

Here we consider third method based on the requirement that the eigenfunction of the third component of the operator of angular momentum must be a single valued periodic function with the period $2\pi$ and analyze whether in this framework it can be proven that the spectrum of angular momentum consists of only integers. Our finding is the same as in \cite{JKT}, namely that again this is not the case - the statement that in the framework of quantum mechanics the 
eigenvalues of angular momentum and its third component are comprised of only integer numbers is not a strictly proven theoretical result.

This article is organized as follows: in section II we discuss  the 
requirements of single valuedness and periodicity of the eigenfunction of the third component of angular momentum. We show that the eigenfunction is a single valued function and note that the requirement of single valuedness and periodicity of wave function has no physical grounding. Therefore establishing the spectrum of observables based on a requirement on a non observable, such as a wave function, is not a self consistent procedure and contradicts to the first principles of quantum mechanics .  The latter was advocated already by W.~Pauli \cite{Pauli}. We discuss the non uniqueness of eigenfunction in spherical and Cartesian coordinates and show that this feature is originated by the non uniqueness related with the operation of rising complex number in an arbitrary (not necessary integer) power.

In  section section III we introduce an alternate definition of power of a complex number which, in distinct of Euler's prescription, $(e^{i\phi})^m=e^{im\phi}$,  retains the invariance with regard translation $\phi\to\phi+2k\pi$, $k$ is integer and $m$ is the eigenvalue of the third component of the operator of angular momentum. In the framework of this new definition the eigenfunctions are the Gauss hypergeometric functions, that are regular, their expansion in series of $\phi$ coincides with the expansion of $\cos(m\phi)$ and $\sin(m\phi)$ and simultaneously they are not equal to $\cos(m\phi)$ and $\sin(m\phi)$. We point out and correct the discrepancy in describing relations between these hypergeometric functions and the trigonometric ones, present in standard references \cite{Bat}, \cite{Abramowitz}.

In section IV we discuss the alternative method of arguments suggested by Pauli \cite{Pauli}. Instead of demanding periodicity of the eigenfunction which he rejected as a requirement on a non observable, Pauli introduced a specific selection rule for matrix elements of angular momentum operator which he considered as a necessary condition for the validity of fundamental commutation relations in matrix form. Based on this selection rule, Pauli derived that eigenvalues of the angular momentum operator could be only integer. Close examination shows that commutation relations held without invoking selection rule. Therefore, first principles of quantum mechanics are satisfied without Pauli's conjecture and we see no unavoidable necessity to retain the latter in the theoretical framework. Consequently, once again there is no proof that theory predicts that the eigenvalues of angular momentum operator are only integer.

%based on which Pauli derived that eigenvalues of the angular momentum operator could be only integer. Since Pauli's requirement 

%Section \ref{secIV} demonstrates a new mathematical basis, alternative to the mathematical basis related to the problem of the third component of the angular momentum.  
In section V we review in the form of conclusions the issues considered in \cite{JKT} and in the present article.\\

\section{On single valuedness of the eigenfunctions of the operator of the third component of angular momentum}
\label{secII}
The third method used to prove that the eigenvalues of the angular momentum and its third component are only integers, is based on the following arguments (see, e.g. \cite{shiff}-\cite{Landau}):
\begin{enumerate}
\item
$\Psi_m(\phi)=\exp(i m\phi)$, the eigenfunction of the operator of the third component of the angular momentum $\hat M_z$, is a non single valued function for a non integer values of $m$.

\item

From the physical point of view it is unacceptable that a wave function corresponding to an observable is a non single valued function.

\item

For $\Psi_m(\phi)=\exp(i m\phi)$ to be a single valued, i.e. to correspond to an observable, $m$ must be an integer.

\end{enumerate}

Let is divide analysis of the above statements in two parts, first, whether $\Psi_m(\phi)=\exp(im\phi)$ is a non single valued function and second, whether wave functions has to be single valued.
%Let us divide in two parts the discussion of the above statement. In the first part we consider the issue if  $\Psi_m(\phi)=\exp(i m\phi)$ is non-single-valued. In the second part we address the question if the quantum-mechanical wave functions have to be single-valued. 

We start discussion of the first point by mentioning that as a rule, the issue of the non single valuedness of wave functions is addressed in spherical coordinates.  
%This is an important detail of the main statement. 
For example, in Ref.~\cite{BL} the eigenfunction equation for the third component of the angular momentum is considered in spherical coordinates
\begin{equation}
(-i\partial/\partial\phi) \Psi_m(\phi) =m \Psi_m(\phi), \  \ \Psi_m(\phi)=\exp(i\,m\phi), 
\label{eq1.1}
\end{equation}
%(We use the system of units where $\hbar=1$) is considered in spherical coordinates and it is stated, 
and without any explanatory comments it is stated that 
"If $\Psi_m(\phi)$ is to be a single-valued function of $\phi$, it is necessary that $m$ should be an integer, $m=\{0, \pm 1, \pm 2, \cdots\}$" (\cite{BL}, Appendix V).
%\begin{equation} m=\{0, \pm 1, \pm 2, \cdots\};\label{eq1.2}\end{equation}
As the text offers no additional comments one may get impression that the function $\exp(i\,m\phi)$ by itself is not single valued. Of course this is not so since to one value of the argument $\phi$ and any fixed $m$ there corresponds a single value 
of the the function $\exp(im\phi)$. The opposite is not true, to a fixed value of $ \Psi_m(\phi)$  correspond numerous values of the argument:
\begin{equation}
\phi(k)=\phi+2\pi k/m,\ \ \ k=0,\pm 1\pm 2,\cdots
\label{eq1.3}
\end{equation}
Discussion of the non single valuedness  of functions $ \Psi_m(\phi) $ is accompanied by the following comment in Ref.~\cite{Landau}:
"to become a single valued function,  $ \Psi_m(\phi) $ must be periodic with the period $2\pi$ and correspondingly the following condition must hold:  $ \Psi_m(\phi)=\Psi_m(\phi+2 \pi k) $, where $k$ is an arbitrary integer number".
If we accept periodicity of a wave function as a physical requirement which must necessarily hold, then indeed $m$ has to be an integer. However, even in this case it has to be emphasized that the condition of periodicity with a period $2\pi$ can not turn $ \Psi_m(\phi) $
into a single valued function, because already without the periodicity requirement $\Psi_m(\phi)$ is a single valued function.

None of these authors indicate explicitly why $\Psi_m(\phi)=\exp(i\,m\phi)$ is considered as a non single valued.  Neither it is explained what physical conditions would be violated if these functions were non single valued. 
The necessity of the condition of periodicity is not elucidated either, therefore, one is left to guess what could have been the underline arguments here.

%considered regarding this issue. 
%Regarding non single valuedness, the source for this feature 
The issue probably stems from the introduction of spherical coordinates. Indeed, as is well known, transition from the Cartesian coordinates to the spherical ones and vice versa is not a one-to-one  correspondence \cite{ww} - to a fixed Cartesian coordinates $(x,y,z)$ correspond fixed spherical coordinates $(r,\theta,\phi)$, however the opposite is not the case. %uniquely fixed Decart coordinates $(x,y,z)$ correspond to a fixed set of the spherical coordinates $(r,\theta,\phi)$, however to fixed Decart coordinates does not correspond a fixed set of spherical coordinates. 
When introducing spherical coordinates, the  periodicity $x(\phi)=x(\phi+2 \pi k)$, $y(\phi)=y(\phi+2 \pi k)$, $z(\phi)=z(\phi+2 \pi k)$ is generated which in turn generates the multi valuedness of $\phi$ given by %and this periodicity is the reason of the non uniqueness of 
$\phi=\arctan(x/y)+2\pi k$. Since the introduction of the spherical coordinates iat the same time is a parameterisation of the physical space, the domain of this variable is restricted to $(0,2\pi)$ where only one of the boundaries should be included as a possible value of $\phi$. The reason is simple: in any coordinate system one physical point should be taken into account only once. Therefore introducing the spherical coordinates does not generate 
the periodicity condition $f(\phi)=f(\phi+2 \pi k)$ for the eigenfunction of the third component of the angular momentum. %On the contrary, according to this mechanism the mentioned property is restricted by the condition that even in case of dynamical rotations after each full turn reading of the coordinate must start from the beginning like the reading of the time in case of mechanical clocks. 
The meaning of  $\phi+2 k \pi$ is not mappping of variable $\phi$ to some process occurring in real space. %not represent mapping ofmapping of some dynamical as the value of argument of the mapping does not exist. There 
The meaning is as follows: there exists variable $\phi$ with the values in range $0\leq \phi < 2 \pi$ or $0< \phi \leq 2 \pi$, and simultaneously there emerges a discrete-valued dynamical quantity - the number of rotations $k$.
\footnote{In many textbooks in quantum mechanics, including already mentioned \cite{shiff}, \cite{BL}, \cite{Landau}
the area of $\phi$ is indicated by a segment closed from both sides, $\phi\in[0,\,2\pi]$. One might think that this is just a typo, or it could be an implicit 
%the appearance of this segment in textbooks of quantum mechanics 
%is to be understood as a potential 
argument for supporting condition of periodicity of $\Psi_m(\phi)$. In any case, this is mathematically inconsistent and incorrect. Correct definition of a domain is of course a segment closed only from one side, $[0, \,2\pi)$ or $(0,\,2\pi]$, see e.g. \cite{ww}, \cite{cop}, \cite{cur}.}

Another argument for the periodicity of the wave function might be related to the following: physical reality observed in a given system of reference should not differ from the physical reality observed in a system 
which has been rotated by $2 \pi$, therefore the corresponding observable quantum mechanical quantities should not differ either. This is again related to the above mentioned non-uniqueness: when we consider two 
reference frames the axis of which coincide there is no operational way to distinguish wether the angle the axis is $0$ or $2\pi k$. %This fact is of course the reason for having non-unique correspondence between Cartesian and spherical coordinates. 
Notice that even if it would be possible to justify the periodicity condition using this argument, it would be still unclear why should one apply periodicity condition to the wave function which is not an observable quantity, see, e.g., Ref.~\cite{Pauli}. 
Clearly, %Moreover, the mechanism of the connection of observable quantities with wave functions guarantees that 
the phase factor related to the multi valuedness,  $\exp(i m 2\pi k)$, which is generated by the rotation, has no effect on the calculated physical quantities. 
One of the axioms of the quantum mechanics states that the wave functions can be specified only modulo numerical phase factor \cite{shiff}, \cite{BL}, \cite{Landau}.
Therefore, from the requirement that physical reality observed in a given system of reference should not differ from the physical reality observed in a system 
which has been rotated by $2 \pi$, it does not follow that the wave function must be periodical with the period $2\pi$.

%neither in this case would be mathematically consistent to connect the condition of periodicity to the physical fact of rotation.

From the mathematical point of view more transparent scheme, avoiding the non-uniqueness generated by introducing spherical coordinates, %the most consistent scheme of argumentation is related to 
is the one analyzing the eigenfunctions of the third component of the angular momentum in Cartesian coordinates:
%The corresponding equation has the form: 
\begin{equation}
\hat M_z \Psi_m (x,y,z) =i(y\, \partial/\partial x-x\, \partial/\partial y)\Psi_m (x,y,z)= m \Psi_m (x,y,z),
\label{eq1.4}
\end{equation}
the solution of which is given as a power function of a complex variable
\begin{equation}
\Psi_m (x,y,z)=f(z; x^2+y^2) (x+i y)^m, 
\label{eq1.5}
\end{equation}
where $f$ is an arbitrary finite differentiable function \cite{Fock}. $\Psi_m$ contains non-uniqueness related with the definition of power function for the case  of the non-integer exponent. %by a non-integer values of the exponent. %The spectrum of this non-uniqueness becomes considerably wider when moving to the space of complex variables. 
This non-uniqueness is not directly related to the non-uniqueness caused by the introduction of spherical coordinates, however the mathematical nature of the both is similar since 
both use the same Euler's identity $\exp(i x)=\cos x+i \sin x$ \cite{ww}-\cite{cur}. Unlike the spherical coordinates, the non-uniqueness related to the power functions shows up for any coordinates - Cartesian, spherical, parabolic etc. Therefore the problem related to this non-uniqueness is purely mathematical. 
To better understand the nature of problems related to the spectrum of the angular momentum, we need to understand details of the mentioned mathematical problem. Consider the rotation of Cartesian coordinates 
in $\phi$ around the $Z$ axis:
\begin{equation}
x = x' \cos\phi -y' \sin\phi; \ \ \ y = x' \sin\phi + y' \cos\phi; \ \ \ z=z' ;
\label{eq1.6}
\end{equation}
The operator $\hat M_z$ should not change due to such rotations. Indeed, using the relations: $(\partial y'/\partial y)=\cos \phi$, $(\partial x'/\partial y)=\sin \phi$, $(\partial y'/\partial x)=-\sin \phi$ and $(\partial x'/\partial x)=\cos \phi$, 
we obtain:
\begin{eqnarray}
\hat M^{\prime}_z (x^{\prime},y^{\prime})\equiv& i(y^{\prime}\, \partial/\partial x^{\prime}-x^{\prime}\partial/\partial y^{\prime})= i(y\, \partial/\partial x-x\, \partial/\partial y) \equiv M_z(x,y)
%\hat M_z &= & i(y\, \partial/\partial x-x\, \partial/\partial y) \nonumber\\
%&=& i\left\{ ( x' \sin\phi + y' \cos\phi) \left[ (\partial x'/\partial x)\partial/\partial x' + (\partial y'/\partial x)\partial/\partial y' \right] \right. \nonumber\\
%&- & \left. (x' \cos\phi -y' \sin\phi) \left[ (\partial x'/\partial y)\partial/\partial xpartial y'/\partial y)\partial/\partial y' \right] \right\} \nonumber\\&=& i(y'\, \partial/\partial x'-x'\, \partial/\partial y'); 
\label{eq1.7}
\end{eqnarray}
Operator $\hat M_z$ possess two symmetries. The first one is related to rotations around $Z$ axis: $\hat M_z(x,y)=\hat M_z(x',y')$. The second is related to scale transformations: $x\rightarrow \lambda x,\;y\rightarrow \lambda y\,\rightarrow\,
\hat M_z(\lambda x, \lambda y)=\hat M_z(x,y)$. 
Not every $\Psi_m$ is rotational and scale invariant. %These symmetries are not automatically transferred to $\Psi_m(x,y)$ functions. However it is possible to satisfy them on the level of solutions. 
For example, if in Eq.~(\ref{eq1.5}) we assume that 
$f(z;x^2+y^2)=1$, then the scale invariance is violated in corresponding $\Psi_m$: $\Psi_m(\lambda x,\lambda y)\neq \Psi_m(x,y)$. Scale invariance holds when the coefficient function is chosen as:
\begin{equation}
f(z;x^2+y^2)=g(z)(x^2+y^2)^{-m/2}; 
\label{eq1.8}
\end{equation}
As for the rotational symmetry, the wave function transforms as follows:
%in this case we should start with the following relation:
\begin{eqnarray}
\Psi_m(x,y,z) & = & f(z;x^2+y^2)(x+i y)^m\,\rightarrow\,f(z;x'^2+y'^2)[(x'+i y')(\cos\phi +i \sin\phi)]^m\nonumber\\
&=& (\cos\phi +i \sin\phi)^m\, \Psi_m(x',y',z');
\label{eq1.9}
\end{eqnarray}
To understand this result, let us consider the obtained relation both from mathematical and from the quantum-mechanical point of view. 

%From the mathematical point of view the construction
Construction $(x+i y)$, unlike $\hat M_z(x,y)$, is non-invariant under the rotations (\ref{eq1.6}), %since in general, 
%Clearly, this has to be like this because the 
%coordinates $x$ and $y$ change under the rotations with angle $\phi$. However, 
however, when $\phi$ is an integer multiple of $2 \pi $, $x+iy$ is invariant, %the condition of invariance is satisfied: 
$(x+i y)\rightarrow(x'+i y')(\cos 2\pi k+i \sin 2\pi k)=(x'+i y')$, in full agreement 
%and this mathematical fact is in a full agreement 
with empirical facts. 
As for the function $\Psi_m(x,y,z)$, due to the rotation on $2\pi k$ angle it acquires numerical factor $(\cos 2\pi k+\sin 2\pi k)^m=(1)^m$. This factor for a non-integer $m$ represents the source of the non-uniqueness related to the 
periodicity in case of power functions. Let us mention once again that this non-uniqueness is related to the non-uniqueness of the power function and is not directly related to periodicity. The fact that $(1)^m$ for non-integer $m$ is not an uniquely defined quantity is 
reflected in the fact that there exist several non-equal quantities $A=\{ A_m\}$ which satisfy the condition $(A_m)^{1/m}$=1. In the spectrum of these quantities there may be ones for which symmetries of the starting expression are satisfied and others for which these symmetries are broken. E.g., let us consider power function $A_m(x)=(x^2)^m$ for $m=1/2$:
\begin{equation}
A_{1/2}(x) = (x^2)^{1/2}= \{ \pm x; \pm |x|\};
\label{eq1.10}
\end{equation} 
The starting function, $A_1=x^2$, is evidently an even function %appearing prior to the square root operation. This function is invariant under the sign-reflection of the argument
: $A_1(x)=A_1(-x)$. 
If we use either of the first two prescriptions of Eq.~(\ref{eq1.10}) for the square root operation - $A_{1/2}^{(1)}=-x$ or $A_{1/2}^{(2)}=x$ then the symmetry characterising the initial function will be spontaneously broken. 
The remaining two prescriptions $A_{1/2}^{(3,4)}=\pm |x|$ retain the original symmetry. Let us consider the power function $B_m(\phi)=(\cos\phi +i \sin \phi )^m$ from the point of view of this phenomenon. 
The starting function of this power function $B_1(\phi)=(\cos\phi +i \sin \phi )$ is invariant under translations with $2\pi k$:
\begin{equation}
B_1(\phi)=B_1(\phi+2\pi k);
\label{eq1.11}
\end{equation}
The prescription, defined by the Euler's identity $\cos x + i\sin x=e^{ix}$
\begin{equation}
B_m(\phi)^{\rm (Euler)}=[B_1(\phi)]^m=(\cos\phi + i\sin\phi)^{m}=[\exp(i\phi)]^m=\exp(i m \phi),
\label{eq1.12}
\end{equation}
which is used in the definition of the power function, leads to the result where the translational symmetry (\ref{eq1.11}) of the starting function is broken for the non-integer $m$:
\begin{equation}
B_m(\phi)^{\rm (Euler)}\neq B_m(\phi+2\pi k)^{\rm (Euler)};
\label{eq1.13}
\end{equation} 
Notice that Euler's prescription (\ref{eq1.12}) is first defined for only integer $m$ (see, e.g. \cite{ww}) and then relation  (\ref{eq1.12}) is extended to a non-integer $m$. 
%We use this prescription as the only correct equality and after based on this we build the complex analysis. 
%The problem connected to the status of the introduction of relation (\ref{eq1.12}) is the following: 
The relation (\ref{eq1.12}) as a mathematical construction that is completely correct and self consistence since it is obtained by using a well-defined mathematical prescription. However this does not mean that there cannot exist 
a different prescription $B_m(\phi)^{\rm (other)}$ that, in distinct of $B_m(\phi)^{\rm(Euler)}$, retains the starting symmetry of Eq.~(\ref{eq1.11}):
\begin{equation}
B_m(\phi)^{\rm (other)}=B_m(\phi+2\pi k)^{\rm (other)}.
\label{eq1.14}
\end{equation} 
One possible example of such a prescription will be given in the next section.  This other prescription for a power function brings in changes in the analysis of the quantum mechanical problem of the angular momentum as well as in the complex analysis.

The technical detail is very similar to one mentioned in Ref.~\cite{JKT}. Let us recapitulate it here. 
%$hat M^2$, the square of the angular momentum satisfies%

The eigenfunctions of $\hat M^2$, the operator of the  square of the angular momentum are given by the solution of the equation% $\hat M^2 \Psi_M(\xi |L;m )= L(L+1)\Psi_M(\xi |L;m )$ 
\begin{equation}
\hat M^2 \Psi_M(\xi |L;m )= L(L+1)\Psi_M(\xi |L;m ),
\label{eq1.15}
\end{equation} 
where $\xi\equiv \cos\theta,\,\theta$ being zenith angle and  the third component of the operator of the angular momentum operator is replaced by its eigenvalue $\hat M_z\to m$. The operator  $\hat M^2$ is an even function of $m$:
\begin{equation}
\hat M^2(m)=\hat M^2(-m),
\label{even}
\end{equation}
%with $m$ being the eigenvalue of the third component of the operator of the angular momentum.
This property of the invariance under the reflection of the sign of $m$ must be present  in the solutions of Eq.~(\ref{eq1.15}). Indeed, as shown in Ref.~\cite{JKT}, Eq.~(\ref{eq1.15}), which is a differential equation of the second order, has two linearly independent solutions 
$\Psi^0_M(\xi |L;m)$ and $\Psi^1_M(\xi | L;m)$ which satisfy the condition:
\begin{equation}
\Psi^0_M(\xi |L;m)= \Psi^0_M(\xi |L; -m); \ \ \ \Psi^1_M(\xi | L;m)=\Psi^1_M(\xi | L; - m); 
\label{eq1.16}
\end{equation} 
These functions themselves are given as products:
\begin{eqnarray}
\Psi^0_M(\xi | L; m)&=& (1-\xi^2)^{\beta} { }_2 F_1 \left( {1\over 2} +\beta+\frac{L}{2},\,\beta-\frac{L}{2},\,\frac{1}{2};\,\xi^2\right), \label{first}\nonumber\\
\Psi^1_M(\xi | L; m)&=& \xi (1-\xi^2)^{\beta} { }_2 F_1 \left( 1+\beta+\frac{L}{2},\,\frac{1}{2}+\beta-\frac{L}{2},\,\frac{3}{2};\,\xi^2\right),
\label{eq1.17}
\end{eqnarray}
where $4\beta^2=m^2$ and ${}_2F_1(a,b;c;\xi)$ are the Gauss's hypergeometric functions. In general the factors in Eq.(\ref{eq1.17}) violate the above mentioned symmetry:
\begin{equation}
\label{gia1}
(1-\xi^2)^{\beta} \neq (1-\xi^2)^{-\beta}; \ \ \ {}_2F_1(a,b;c;\xi)|_{\beta}\neq {}_2F_1(a,b;c;\xi)|_{-\beta}.
\end{equation}
That is, when representing solutions to the Eq.~(\ref{eq1.15}) in the form of Eq.~(\ref{eq1.17}) we are moving from the class of invariant functions $\Psi_M(\xi | L;m)$ to the class of non-invariant functions. 
This transition is determined by the definition of the square root in $2\beta=(m^2)^{1/2}$ and thus is completely spontaneous. If we use the prescription  $(m^2)^{1/2}=\pm |m|$, then 
the factors in (\ref{eq1.17}) are individually invariant  at $m\to -m$. If we choose the prescription $(m^2)^{1/2}=\pm m$ then the factors in Eq.~(\ref{eq1.17}) are extended to the class of non-invariant functions. The products, $\Psi^{0,\,1}_M(\xi | L;m)$, are of course invariant for any prescription for $(m^2)^{1/2}$.
Prescriptions  $(m^2)^{1/2}=\pm |m|$ and $(m^2)^{1/2}=\pm m$ lead to quite different solutions to Eq.~(\ref{eq1.15}): in case of realising the conditions of polynomialization, imposed on solutions to avoid singularities (see Ref.~\cite{JKT}) for $(m^2)^{1/2}=\pm |m|$ the sets of eigenfunctions consist of normalisable function only, and in case of $(m^2)^{1/2}=\pm m$ they contain normalisable as well as non normalisable functions. 

Similarly, when analyzing the eigenvalue/eigenfunction problem for the third component of the angular momentum, one may anticipate that there exists prescription for the power function %Analogously, we may guess that when indicating the eigenfunctions of the third component of the angular momentum for the operation of taking in power one can find such a prescription 
for which the function $B_m(\phi)=(\cos\phi+i\sin\phi)^m$ 
has the symmetry (\ref{eq1.11}) of the starting expression and the Eq.~(\ref{eq1.14}) is satisfied. In this case the eigenfunction of the third component will be defined in such a way that the problem on non-uniqueness, related to the 
rotation, will not appear for the power functions at all. Within such a prescription the following relation must hold:
\begin{equation}
\label{gia2}
B_m(2 \pi k)=(\cos 2\pi k+i\sin 2\pi k)^m=(1)^m=1.
\end{equation}
We finalise the discussion of mathematical aspects of Eq.~(\ref{eq1.9}) by pointing out one more feature. To do so we present the eigenfunction equation for $\hat M_z$ in Cartesian and spherical coordinates:
\begin{eqnarray}
\label{gia3}
\hat M_z \Psi_m(x,y)&=&i(y\partial /\partial x - x\partial /\partial y )\Psi_m(x,y)=m \Psi_m(x,y);\\
%\end{equation}
\hat M_z \Psi_m(\phi)&=&- i (\partial /\partial \phi) \Psi_m(\phi)=m \Psi_m(\phi).
\end{eqnarray}
Equation in the spherical coordinates under the scale transformation $\phi'=m \phi$ satisfies following condition:
\begin{equation}
- i (\partial /\partial \phi') \Psi_m(\phi'/m)=\Psi_m(\phi'/m).
\label{eq1.18}
\end{equation}
Nothing similar to (\ref{eq1.18}) is observed in Cartesian coordinates. This means that transiting from Cartesian to spherical coordinates generates properties which are absent in Cartesian coordinates. This is what defines the difference between eigenfunctions of the third component of the angular momentum in two coordinate systems. In particular, in Cartesian coordinates the eigenfunctions of $\hat M_z$ belong to the class 
of power functions of the coordinates. In spherical coordinates, using the Euler's prescription, the eigenfunctions belong to the class of exponential functions of the spherical coordinate - not to the class of the power functions. 

Let us discuss Eq.~(\ref{eq1.9}) from the quantum mechanical point of view. 

Under rotations of the reference frame on angle $2 \pi k$ the operator $\hat M_z$ does not change while the corresponding eigenfunctions 
gain a factor which for the Euler's prescription has the form:
\begin{equation}
c_m=(\cos 2\pi k + i \sin2\pi k)^m= 
\left\{
\begin{array}{c}
1^m \\
\exp (i 2\pi k m) 
\end{array}
\right. .
\label{eq1.20}
\end{equation}
Remember that in quantum mechanics wave function is defined up to a $c$-number numerical factor which allows to introduce the phenomenon of normalised wave functions and after normalization is performed the phase of the $c$-number factor 
still remains arbitrary \cite{shiff}-\cite{Landau}. The quantity in Eq.~(\ref{eq1.20}) fits within this non-uniqueness of the phase factor and from the quantum mechanical point of view it makes little, if any  sense to try to connect it with any physical phenomenon. Clearly the requirement of single valuedness of wave function is not supported by any physical principle. Exactly this point of view is stated by Pauli in \cite{Pauli}, to which we completely agree and the details
of Pauli's statement will be discussed below in section \ref{secIII}.

\section{On alternative definition of the power function}
\label{secIV}

Consider a representation of the power function: 
\begin{equation}
\Psi(m;\phi)=(\cos \phi+ i \sin \phi)^m=\exp(i\,m\phi),
\label{eq3.1}
\end{equation}
which is obtained by exponentiating Euler's identity $\cos \phi+ i \sin \phi =\exp(i\,\phi)$. 
This identity has the following property of translational invariance:
\begin{eqnarray}
\cos \phi+ i \sin \phi &=& \cos (\phi+2 \pi k)+ i \sin (\phi+2 \pi k) ,\nonumber\\
\exp(i\,\phi) & = & \exp(i\,\phi+i \,2 \pi k),
\label{eq3.2}
\end{eqnarray}
which is spontaneusly broken in Eq.~(\ref{eq3.1}). As a result of this breaking the following numerical relation emerges:
\begin{equation}
[\cos (2 \pi k)+ i \sin (2\pi k)]^m=(1)^m=\cos (2 \pi k m)+ i \sin (2\pi k m).
\label{eq3.3}
\end{equation}
This leads to the following spectrum 
\begin{equation}
(1)^m=\{ 1; \cos (2 \pi m)+ i \sin (2\pi m); \cos (4 \pi m)+ i \sin (4\pi m);\cdots \};
\label{eq3.4}
\end{equation}
For $m$ integer invariance regarding translations $\phi\to\phi+2k\pi$ is restored in Eq.~(\ref{eq3.1}) and  the spectrum (\ref{eq3.4}) collapses into $(1)^m=\{1;1;1;\cdots\}=1$. %For integer values of $m$ the translational invariance of Eq.~(\ref{eq3.1}) is restored. 
For the non-integer values of $m$ Euler's representation of power function has a simple problem. Namely, in the algebra of real numbers as well as in the complex analysis the following statement the following axiom is implied: 
if any mathematical expression $X$ is multiplied by unity, this expression should not change. Algebraically this axiom is written as $X=1\times X$. Another axiom is expressed by the following relation:%It is also implied that
%It is also an axiom the statement expressed by the following relation:
\begin{equation}
F(X)=F(1\times X)=F(1\times 1\times X) = \cdots = F((1)^k \times X) = \cdots ,
\label{eq3.5}
\end{equation}
where $F(X)$ is some function of $X$. 
%According to the corresponding axiom 
Eq.~(\ref{eq3.5}) is postulated to be an identity and applying it to the 
%Using the mentioned identity in 
Euler's representation of the power function we obtain
\begin{equation}
(\cos \phi+ i \sin \phi)^m =[(1)^k(\cos \phi+ i \sin \phi)]^m =(1)^{m k} \exp (i m \phi ) ,
\label{eq3.6}
\end{equation}
%For non-integer $m$
For all the values of the spectrum of Eq.~(\ref{eq3.4}) and non-integer $m$ Eq.~(\ref{eq3.6}) is no longer an identity, moreover, it yields wrong results.
%is not only an identity, but rather for non-integer $m$ leads to incorrect equalities. 
The reason for this  is known and is related to the non-uniqueness of 
 power function with the non integer exponents \cite{cop}, \cite{cur}. One of the main characteristics of this non-uniqueness is the multiplicity of the spectrum of Eq.~(\ref{eq3.4}). For example, in case of $m=1/2$ we have:
\begin{equation}
(1)^{1/2}=\{1;-1;1;-1; \cdots \};
\label{eq3.7}
\end{equation}
This set contains only two numerical values which can be indicated through the following simple relation in the algebra of real numbers: $(1)^{1/2}=\left\{ \begin{array}{c} +1\\ -1 
\end{array} \right.$. The same relation can be also written as $(1)^{1/2}=\left\{ \begin{array}{c} -1\\ +1 
\end{array} \right.$. That is, the ordering of elements plays no role in indicating the relation between the left hand and right hand sides of the equality: $(\pm 1)^2=[(1)^{1/2}]^2=1$. 
The form of Eq.~(\ref{eq3.7}) is a concrete presentation of the elements in this set, which is known in mathematics as introduction of the ordering \cite{rud}. 

E.g., when using the ordering of the form of Eq.~(\ref{eq3.7}) in the theory of complex variables, for the square root we obtain:
\begin{eqnarray}
(\cos \psi+ i \sin \psi)^{1/2} &=& [\cos (\phi+2 \pi k)+ i \sin (\phi+2 \pi k)]^{1/2} = \exp(i\,\psi/2)= \exp(i\,\phi/2 +i\pi k); \nonumber\\
\psi & = & \phi + 2 \pi k, \ \ \ 0\leq\phi < 2\,\pi.
\label{eq3.8}
\end{eqnarray}
Eq.~(\ref{eq3.8}) is interpreted as follows: the result is a function with two sheets in which one sheet corresponds to even values of $k$ and another to odd values. That is, if the numerical values of $\phi $ are located in $[0, 2\pi ],[4\pi,6\pi ],\cdots $ segments we get the periodically arranged 
set corresponding 
to the first sheet and if the numerical values of $\phi$ are located in $[2\pi,4\pi], [6\pi,8\pi ],\cdots $ segments we get the periodically arranged set corresponding 
to the second sheet \cite{ww}-\cite{cur}. In other words, Euler's prescription for the power of a complex number, Eq.~(\ref{eq1.12}),  implies the realisation of all possible values in the spectrum emerging in taking power of a complex number and an introduction of 
a certain ordering in this set of values is understood. Therefore one can say that in the theory of complex variables Eq.~(\ref{eq1.12})  is a concrete prescription for resolving the non-uniqueness related to the operation of taking power 
of a complex number by introducing specific ordering of a set of values of a power. Evidently, Euler's prescription (\ref{eq1.11}) violates the property (\ref{eq3.2}) of translational invariance of the starting expression. 

One may wonder if it is possible define function  $\Psi(m;\phi)$ so that the 
the translational invariance of the starting expression is preserved, i.e. that for any $m$ we have
\begin{equation}
\Psi(m;\phi)=\Psi(m; \phi+2 \pi k).
\label{eq3.9}
\end{equation}
To address this issue let us rewrite Eq.~(\ref{eq1.1}) written for complex functions in the form of equations for real and imaginary parts of $\Psi(m;\phi)$:
\begin{eqnarray}
{\rm d} \Psi_R/{\rm d}\phi & = & - m \Psi_I , \ \ {\rm d}\Psi_I/{\rm d}\phi = m \Psi_R, \nonumber\\
{\rm d}^2 \Psi_R/{\rm d}\phi^2 & = & - m^2 \Psi_R , \ \ {\rm d}^2 \Psi_I/{\rm d}\phi^2 = -m^2 \Psi_I\,.
\label{eq2.10n}
\end{eqnarray}
These second order differential equations are have two linearly independent solutions.

Simplest realization of these solutions is given by $\Psi^{Euler}_R=\cos(m \phi)$ and $\Psi^{Euler}_I=\sin(m \phi)$, which correspond to the complex function given by Euler's parameterization of Eq.~(\ref{eq3.1}).

Different solutions are obtained if a substitution of variables $z=\sin^2\phi$ is made in Eq.~(\ref{eq2.10n}). 
In this case, analogously to the case with the square of the angular momentum the equation for Gaus's hypergeometric function is obtained (see, e.g. Ref.~\cite{JKT} or Ref.~\cite{Abramowitz}):
\begin{eqnarray}
&& \left[ z(1-z) {\rm d}^2/{\rm d} z^2 + [c-(a+b+1) z] {\rm d}/{\rm d} z-a b 
\right] \Psi _{R,I} =0;  \nonumber\\
&& c=1/2; \ a=m/2; \ b=-m/2.
\label{eq2.11n}
\end{eqnarray}
%is obtained (see, e.g. Ref.~\cite{JKT} or Ref.~\cite{Abramowitz}).
This equation has the following solutions:
\begin{eqnarray}
\Psi^0 &=& {}_2F_1\left( a,b;c;z \right) = {}_2F_1\left( m/2,-m/2;1/2; \sin^2 \phi \right) ;  \nonumber\\
\Psi^1&=& z^{1-c} {}_2F_1\left( a-c+1,b-c+1;2-c;z \right) \nonumber\\
&=& \sin\phi \, {}_2F_1\left( m/2+1/2, - m/2+1/2;3/2; \sin^2 \phi \right) ; 
\label{eq2.12n}
\end{eqnarray}
Note that since for the given values of $a$, $b$ and $c$ Eq.~(\ref{eq2.11n}) is invariant under $z\to 1-z$, instead of linearly independent solutions of Eq.~(\ref{eq2.12n}) we can use two alternative solutions specified as:
\begin{eqnarray}
\Psi^{00} &=& {}_2F_1\left( m/2,-m/2;1/2; \cos^2 \phi \right) ;  \nonumber\\
\Psi^{11}&=& \cos\phi \, {}_2F_1\left( m/2+1/2, - m/2+1/2;3/2; \cos^2 \phi \right) ; 
\label{eq2.13n}
\end{eqnarray}
Using these solutions, functions  $\Psi_R$ and $\Psi_I$, satisfying equations (\ref{eq2.10n}) should be constructed. %so that equations (\ref{eq2.10n}) are satisfied. 
For example, we can choose $\Psi_R=\Psi^0$. The corresponding function $\Psi_I$, imaginary part of $\Psi$, is given by the following expression: 
\begin{eqnarray}
\Psi_I &=& -(1/m) {\rm d}\Psi_R/{\rm d}\phi = -(1/m) {\rm d}\Psi^0/{\rm d}\phi = m \cos\phi \sin\phi \, {}_2F_1\left( 1+m/2,1-m/2;3/2; \sin^2 \phi \right)  \nonumber\\
&=& m \sin\phi \, {}_2F_1\left( m/2+1/2, - m/2+1/2;3/2; \sin^2 \phi \right) = m \Psi^1 ; 
\label{eq2.14n}
\end{eqnarray}
This $\Psi_I$ indeed satisfies Eq.~(\ref{eq2.10n}):
\begin{eqnarray}
{\rm d}\Psi_I/{\rm d}\phi &=& m \, {\rm d}\Psi^1/{\rm d}\phi = (m/2 \sin\phi) 2 \cos\phi \sin\phi \, {}_2F_1\left( m/2+1/2,1-m/2+1/2;1/2; \sin^2 \phi \right)  \nonumber\\
&=& m \, {}_2F_1\left( m/2, - m/2;1/2; \sin^2 \phi \right) = m \Psi_R ; 
\label{eq2.15n}
\end{eqnarray}
One can also construct the $\Psi_R$ and $\Psi_I$ functions for the pair of solutions in Eq.~(\ref{eq2.13n}). One can also obtain other pairs of solutions corresponding to 
other substitutions of variables. It is not our aim to enumerate such pairs. Therefore as a demonstration we quote only one possible pair: 
\begin{eqnarray}
t &=&(1-\sin\phi )/2;\nonumber\\
&& \hspace{-1cm} [t(1-t)({\rm d}/{\rm d} t)^2 + (1/2-t)({\rm d}/{\rm d} t) +m^2]\Psi=0; \, c=1/2; a=m; b=-m; \nonumber\\
\Psi^0 & =& {}_2F_1\left( a,b;c;t \right) = {}_2F_1\left( m/2,-m/2;12; (1-\sin\phi)/2 \right) ;  \nonumber\\
\Psi^1 & = & f^{1-c} {}_2F_1\left( a-c+1,b-c+1;2-c;t \right) \nonumber\\
&=& [(1-\sin\phi)/2]^{1/2} \, {}_2F_1\left( m+1/2, - m+1/2;3/2; (1-\sin \phi)/2 \right).
\label{missed}
\end{eqnarray}

We are interested only in whether the functions $\Psi^{Hyperg}_R$ and $\Psi^{Hyperg}_I$ obtained using the above given hypergeometric series satisfy the condition of translational invariance of Eq.~(\ref{eq3.9}):
\begin{equation}
\Psi^{Hyperg}_R(\phi +2\pi k) = \Psi^{Hyperg}_R(\phi);  \ \
\Psi^{Hyperg}_I(\phi +2\pi k) = \Psi^{Hyperg}_I(\phi).
\label{eq2.16n}
\end{equation}
Since the arguments of the hypergeometric functions are periodic functions of $\phi$, translational invariance is satisfied. 

A natural question arises, what is the connection between the functions $\Psi_R$ and $\Psi_I$ and the functions obtained using the Euler's prescription: 
$\Psi^{Euler}_R=\cos(m \phi)$ and $\Psi^{Euler}_I=\sin(m \phi)$? 
First of all let us enumerate the similarities:

i) Both  $\Psi^{Hyperg}$ and $\Psi^{Euler}$ are  infinitely differentiable, continuous and single-valued.\\

\noindent This statement is trivial for the functions $\Psi^{Euler}_R=\cos(m\phi)$ and $\Psi^{Euler}_I=\sin(m\phi)$. It is also not difficult to prove this property for $\Psi^{Hyperg}_R$ and $\Psi^{Hyperg}_I$. Indeed, using Eq.~(\ref{eq2.10n}) we readily obtain that the derivatives exist and they are finite:%can write relations which are automatically satisfied for these functions:
\begin{eqnarray}
\frac{{\rm d}^{2k}}{{\rm d}\phi^{2k}} \Psi^{Hyperg}_R &=& (-m^2)^k\Psi^{Hyperg}_R,\quad \frac{{\rm d}^{2k+1}}{{\rm d}\phi^{2k+1} } \Psi^{Hyperg}_R= -m (-m^2)^k\Psi^{Hyperg}_I, \nonumber\\
\frac{{\rm d}^{2k}}{{\rm d}\phi^{2k}} \Psi^{Hyperg}_I&=& (-m^2)^k\Psi^{Hyperg}_I,\quad\frac{{\rm d}^{2k+1}}{{\rm d}\phi^{2k+1} } \Psi^{Hyperg}_I =-m (-m^2)^k\Psi^{Hyperg}_R.
\label{eq2.17n}
\end{eqnarray}
By taking into account that the functions $\Psi^{Hyperg}$ %corresponding to Eq.~(\ref{eq2.13n}) 
satisfy conditions $\Psi^{Hyperg}_R(m;\phi) = \Psi^{Hyperg}_R(m; -\phi) $ and $\Psi^{Hyperg}_I(m;\phi) = -\Psi^{Hyperg}_I(m; -\phi)$ it is straightforward to show that the series expansion of 
these functions is as follows:%can be expanded in powers of $\phi$ as follows:
\begin{eqnarray}
\Psi^{Hyperg}_R(m; \phi) & = & \Psi^{Hyperg (k)}_R(m; 0)\,{\phi^k\over k!}= (-m^2)^k \,{\phi^{2k}\over (2 k)!},  \nonumber\\
\Psi^{Hyperg}_I(m; \phi) & = & \Psi^{Hyperg (k)}_I(m; 0)\,{\phi^k\over k!} =m (-m^2)^k {\phi^{2k+1}\over (2 k+1)!}.
\label{eq2.18n}
\end{eqnarray}

Therefore, the second main similarity of functions $\Psi^{Hyperg}$ and $\psi^{Euler}$ is:

ii) Taylor expansions of the functions $\Psi^{Hyperg}_R(m;\phi)$ and $\Psi^{Euler}_R(m;\phi)=\cos(m\phi)$ coincide; Taylor expansions of the functions $\Psi^{Hyperg}_I(m;\phi)$ and $\Psi^{Euler}_I(m;\phi)=\sin(m\phi)$ also coincide.

Despite sharing properties i) and ii) %there is a substantial difference between 
the functions $\Psi^{Hyperg}$ and $\Psi^{Euler}$ differ substantially. This difference is the reason why the relations given in some standard references 
%which does not allow to write those equalities which are given in some mathematical handbooks (
(see e.g. Ref.~\cite{Abramowitz}, {\bf 15.1.15,\,15.1.17})
\begin{eqnarray}
&& {}_2F_1(-m/2,m/2;1/2;\sin^2\phi) =\cos(m\phi); %\ ({\rm see} \ %\cite{Abramowitz}-15.1.17) 
\nonumber\\
&& m\,\sin\phi \, {}_2F_1(1/2-m/2,1/2+m/2;3/2;\sin^2\phi) =\sin(m\phi) %\ \ ({\rm see} \ %\cite{Abramowitz}-15.1.15)
\label{eq2.19n}
\end{eqnarray}
are Incorrect. Indeed, as we seen above, $\Psi^{Hyperg}$ satisfy the condition of translational invariance (\ref{eq2.16n}) and the functions $\Psi^{Euler}$ do not satisfy them. This means that these functions are not equal. This is very interesting phenomenon 
which needs to be studied however it is beyond the scope of this paper. The only comment we make here is that this the necessity of restricting Eq.~(\ref{eq2.19n}) has been also noticed by some mathematicians \cite{nist}. For us it is important that we have been able to construct such an eigenfunctions of the operator of the third component of the angular momentum, that, unlike the functions $\Psi^{Euler}=\{\cos(m\phi),\,\sin(m\phi)\}$ satisfy the condition of periodicity of (\ref{eq2.16n}). 

To summarize, prescription for a power of a complex number, retaining translational invariance for any real exponent $m$ is as follows:
\begin{equation}
\label{pres1}
(x+iy)^m = \rho^m\,[\Psi_R(\alpha,\,m\,\phi)\,+\,i\Psi_I(\beta,\,m\,\phi)]
\end{equation}
where $\Psi_{R,\,I}$ are solutions of Eq.~(\ref{eq1.1}) presented in terms of functions with arguments periodic in $\phi$,  $\rho$ is an arithmetic root, $\rho=+\sqrt{x^2 + y^2}$, $\phi=\tan^{-1}(y/x)$ and $\alpha, \beta$ are parameters appearing in the solutions. 

Two possible $\Psi_{R,\,I}$ were given  above; let us quote one concrete possibility:
\begin{eqnarray}
\label{pres2}
(\cos\phi +i\sin\phi)^m & = & {}_2F_1\left( m/2,-m/2;1/2; \sin^2 \phi \right)+\nonumber\\
&& i\, m \sin\phi \, {}_2F_1\left( m/2+1/2, - m/2+1/2;3/2; \sin^2 \phi \right).
\end{eqnarray}
For $m$ integer (\ref{pres1}) turns into the Euler prescription $(x+iy)^m=\rho^m\,(\cos(m\phi)+i\sin(m\phi))$ which is invariant for $\phi\to\phi+2k\pi$.

Lastly, let us discuss the hermiticity of the operator $\hat M_z$ with regarding solutions.% $\Psi^{Hyperg}$, let us discuss hermiticity of the operator $\hat M_z$. 
%The condition is:% be Hermitian which means that the following must hold:
Operator is Hermitian if the following relation is satisfied:
\begin{equation}
\label{eq2.20n}
\int_0^{2\pi} d\phi \,\Psi(m';\phi)^* \hat M_z\Psi(m;\phi) = \int_0^{2\pi} d\phi \,[ \hat M_z \Psi(m';\phi)]^* \Psi(m;\phi).
\label{eq2.20n}
\end{equation}
For the condition of hermiticity to hold, the surface term in
\begin{eqnarray}
\int_0^{2\pi} d\phi \,\Psi(m';\phi)^* (-i \partial/\partial \phi )\Psi(m;\phi) &=& -i [ \Psi(m';2\pi)^* \Psi(m;2\pi) - \Psi(m'; 0)^* \Psi(m; 0) ] \nonumber\\
&+& \int_0^{2\pi} d\phi \,[ -i \partial/\partial\phi \, \Psi(m';\phi)]^* \Psi(m;\phi) ; 
\label{eq2.21n}
\end{eqnarray}
must vanish. If in Eq.~(\ref{eq2.21n}) we use functions of $\Psi^{Euler}=\exp(im\phi)$, for the surface term we obtain:
\begin{equation}
i\left(1-e^{i2\pi(m-m')}\right)
%\frac{ i m \left\{ e^{i(m-m') 2\pi }-1\right\}}{m-m'} &=& -i \left\{ e^{i(m-m') 2\pi} b-1\right\} - \frac{ i m' \left\{ e^{ i(m-m') 2\pi} -1\right\}}{m-m'} ; 
\label{eq2.22n}
\end{equation}
Expression (\ref{eq2.22n}) vanishes when $m-m'$ is an integer, i.e. for the operator $\hat M_z$ to be Hermitian, it is not necessary that eigenvalues of $\hat M_z$ are integer; the spectrum with the $m-m'$ integer is sufficient.
%For this equation to become an identity its is necessary that $m-m'$ is an integer number. 
In this case the  orthogonal basis can be constructed using functions  $\Psi^{Euler}$. When using functions  $\Psi^{Hyperg}(m;\phi) = \Psi^{Hyperg}_R(m;\phi) + i \Psi^{Hyperg}_I (m;\phi)$, the surface term $\Psi(m';2\pi)^* \Psi(m;2\pi) - \Psi(m'; 0)^* \Psi(m; 0)$  vanishes for any values of $m$ and $m'$ since according to (\ref{eq2.16n}) $\Psi^{Hyperg}(m;0)=\Psi^{Hyperg}(m;2\pi)$, therefore hermiticity is respected.
%This is 
%guaranteed by relations of Eq.~(\ref{eq2.16n}) - 
%From this point of view the case of functions $\Psi^H(m;\phi)$ is similar to the one which we got when analyzing the spectrum of eigenvalues and eigenfunctions of the square of the angular momentum in Ref.~\cite{JKT}. 

\section{On Pauli's approach to quantization of the angular momentum}
\label{secIII}

Pauli criticised requirement of single valuedness of  wave functions and instead introduced a new condition \cite{Pauli} (see also \cite{merz}). %To make the essence of the Pauli's considerations completely clear let 

Let us quote from his work: 
{\it "As I mentioned in one of my previous papers, there is no {\it a priori} convincing argument stating that the wave functions which describe some physical states must be single valued functions. For physical quantities, which are expressed by squares of wave functions, to be single valued it is quite sufficient that after moving around a closed contour these functions gain a factor $\exp(i\alpha)$, ... In addition let us introduce the following physical condition: 
by applying components 
%by acting with the components 
of the angular momentum $\hat M_k$ ($k=x,y,z$) to 
a given system of regular (i.e. quadratically integrable) eigenfunctions of the square of angular operator $\hat M^2$ corresponding to a fixed eigenvalue we should not leave this system of eigenfunctions ... Of course it is implicitly implied in our condition that the functions obtained by acting 
with $\hat M_k$ must be also regular."} \cite{Pauli}.

As already mentioned above in section \ref{secII} we completely agree with Pauli that the requirement of the single valuedness of the wave function is not physical. Regarding the new condition imposed by Pauli as an alternative to the above requirement of single valuedness of the wave function we state that 
{\it from the physical point of view condition introduced by Pauli is by no means better than the requirement of single valuedness of the wave function declared by himself as unsatisfactory.} To better understand why it is so %the essence of this statement 
let us follow the logic of arguments by Pauli. 

The cornerstone of arguments is as follows: for a non integer $L$ and $m$ it is possible to construct the eigenfunctions of $\hat M^2$ and $\hat M_z$ such that the relations 
\begin{equation}
\label{paul1}
\langle L,m|\hat M_x|L', m\rangle =0,\quad \langle L,m|\hat M_y|L', m\rangle =0
\end{equation}
will be violated, and (\ref{paul1}) should always hold providing $L\neq L'$ \cite{Pauli}. According to Pauli, Eq.~(\ref{paul1}) is nothing else but a special case of general selection rules for the matrix elements of commuting self adjoint operators. Let us consider these rules in details.

As is well-known, to construct a Hilbert space, we should start by identifying the complete set of commuting operators in $\{\hat M^2,\,\hat M_x,\,\hat M_y,\,M_z\}$ (see e.g. \cite{shiff}).  Without the loss of generality one can chose the set $\{\hat M^2,\,\hat M_z\}$ and as a result, the status of the operators $\hat M_z$ and $\hat M_x,\,\hat M_y$ differ.  Indeed, since the Hilbert space is spanned by the eigenfunctions of $\hat M^2$ and $\hat M_z$, the selection rules for matrix elements of these operators are
\begin{eqnarray}
\label{paul2}
\langle L,m|\hat M_z|L',m'\rangle =m\,\delta_{LL'}\delta_{mm'},\quad \langle L,m|\hat M^2|L',m'\rangle =L(L+1)\,\delta_{LL'}\delta_{mm'}.
\end{eqnarray}
Evidently, in the Hilbert space spanned by eigenfunctions of $\hat M^2$ and $\hat M_z$ the matrices  of functions of these operators given by series 
\begin{equation}
f(\hat M^2;\hat M_z)=\sum_{k,l=0}^\infty b_{k l} (\hat M^2)^k (\hat M_z)^l 
\label{eq2.1}
\end{equation}
are also diagonal.

To demonstrate that the status of operators $\hat M_x$ and $\hat M_y$ is different let us discuss how the selection rules (\ref{paul1}) are derived. This is done by considering relations $[\hat M^2,\,M_x]=0$ and $[\hat M^2,\,M_y]=0$ sandwiched by $|L,m\rangle$ and $|L',m\rangle$.  Formal derivation is as follows:
\begin{eqnarray}
\langle L,m|\left[\hat M^2, \hat M_x \right]|L',m\rangle &=& %\langle L,m|\hat M^2 \hat M_x- \hat M_x \hat M^2 |L',m\rangle = (
\langle L,m|0|L',m\rangle = \langle L,m|\hat M^2\hat M_x -\hat M_x\hat M^2|L',m\rangle =\nonumber\\
&& (L-L')(L+L'+1)\langle L,m|\hat M_x|L',m\rangle = 0, \nonumber \\
\langle L,m|\left[ \hat M^2,\hat M_y \right] L',m\rangle &=&%\langle L| \hat M^2 \hat M_y- \hat M_y \hat M^2 |L' \rangle = (
\langle L,m|0|L',m\rangle = \langle L,m|\hat M^2\hat M_y -\hat M_y\hat M^2|L',m\rangle =\nonumber\\
&&(L-L')(L+L'+1)\langle L,m|\hat M_y|L',m\rangle = 0.
\label{defPs}
\end{eqnarray}
Obviously, when $L\neq L'$, relations (\ref{defPs}) are satisfied if  (\ref{paul1}) is valid. Let us recall that the Pauli's suggestion is exactly that the selection rules (\ref{paul1}) are physical, and not the formal requirements of theory. The logic is that the operators of observables commute, $\left[\hat M^2, \hat M_{x,\,y}\right]=0$, so the vanishing of the matrix elements of the commutator is a physical requirement that theory should satisfy and this fact can not be demonstrated unless the selection rules (\ref{paul1}) are valid. %Indeed, if (\ref{paul1}) is violated, from (\ref{defPs}) it will follow that the validity of the cornerstone of theory, $\left[\hat M^2, \hat M_{x,\,y}\right]=0$, can not be demonstrated. 
But, to satisfy the fundamental selection rule $\langle L,m|\left[\hat M^2, \hat M_{x,\,y}\right]|L',m\rangle=0$, we don't necessarily need (\ref{paul1}) and thus the status of the selection rule (\ref{paul1}) is not necessarily a physical requirement.

To show that the vanishing of   $[\hat M^2,\,M_x]$  can be verified without (\ref{paul1}), we need to use relations derived in \cite{JKT}:
\begin{eqnarray}
%hat M^\pm &=& \hat M_x \pm i \,\hat M_y ; \ \ \hat M_x = (\hat M^+ + \hat M^-)/2 ; \ \ \hat M_y = (\hat M^+ - \hat M^-)/(2\,i) ; \nonumber\\
\hat M^- \Psi^0_{Mm}(\phi;\xi|L;m) & = & -(m+L)(L-m+1) \Psi^1_{Mm}(\phi;\xi|L;m-1) ,\nonumber\\
\hat M^- \Psi^1_{Mm}(\phi;\xi|L;m) & = & \Psi^0_{Mm}(\phi;\xi|L;m-1) ,\nonumber\\
\hat M^+ \Psi^0_{Mm}(\phi;\xi|L;m) & = & (L-m)(L+m+1) \Psi^1_{Mm}(\phi;\xi|L;m+1) ,\nonumber\\
\hat M^+ \Psi^1_{Mm}(\phi;\xi|L;m) & = & -\Psi^0_{Mm}(\phi;\xi|L;m+1) ,\nonumber\\
\Psi^{0,\,1}_{Mm}(\phi;\xi|L;m) & = & e^{i m\phi} \Psi^{0,\,1}_{M}(\xi|L;m) , \nonumber\\
\hat M^2(m^2) \Psi^{0,\,1}_{Mm}(\phi;\xi|L;m) & = & L(L+1) \Psi^{0,\,1}_{Mm}(\phi;\xi|L;m).
%\Psi^1_{Mm}(\phi;\xi|L;m) & = & e^{i m\phi} \Psi^1_{M}(\xi|L;m) ; \nonumber\\
%hat M^2(m^2) \Psi^1_{Mm}(\phi;\xi|L;m) & = & L(L+1) \Psi^1_{Mm}(\phi;\xi|L;m) ; 
\label{paul3}
\end{eqnarray}
In (\ref{paul3})  $\hat M^\pm\equiv M_x\pm iM_y$, $\Psi^{0,\,1}_{Mm}$ 
are eigenfunctions of $\hat M^2$ and Eqs.~(\ref{paul3}) are valid for the regular, denoted by $\Psi^{0R,\,1R}_{Mm}(\phi;\xi|L;m)$, as well as for the singular 
$\Psi^{0S,\,1S}_{Mm}(\phi;\xi|L;m)$
%$\Psi^{0S,\1S}_{Mm}(\phi;\xi|L;m)$ 
eigenfunctions of $\hat M^2$.

As an example let us consider the action of the operator $[\hat M^2,\hat M_x]$ on functions $\Psi^0_{Mm}(\phi;\xi| L;m)$:
\begin{eqnarray}
&& [\hat M^2,\hat M_x] \Psi^0_{Mm}(L;m) = (\hat M^2 \hat M_x- \hat M_x \hat M^2 ) \Psi^0_{Mm}(L;m) \nonumber\\
&&= \{\hat M^2 (\hat M^+ +\hat M^- )/2- L(L+1)\hat M_x \} \Psi^0_{Mm}(L;m) \nonumber\\
&&= \hat M^2 \{ (L-m )(L+m+1) \Psi^1_{Mm}(L;m+1) 
- (m+L)(L-m+1) \Psi^1_{Mm}(L;m-1)\}/2 \nonumber\\ 
&&
-L(L+1) \hat M_x \Psi^0_{Mm}(L;m) \nonumber\\
&& = L(L+1) \{ (L-m )(L+m+1) \Psi^1_{Mm}(L;m+1) 
- (m+L)(L-m+1) \Psi^1_{Mm}(L;m-1)\}/2 \nonumber\\ 
&& - \hat M_x \Psi^0_{Mm}(L;m) = L(L+1) \{ \hat M_x \Psi^0_{Mm}(L;m) - \hat M_x \Psi^0_{Mm}(L;m) \} = 0.
\label{2.2N}
\end{eqnarray}
Thus, when acting with the commutator $[\hat M^2, \hat M_x]$ on eigenfunctions of $\hat M^2$ and $\hat M_z$, fundamental relation $[\hat M^2, \hat M_x]=0$ is satisfied without requiring sum rule (\ref{paul1}) to hold.
Therefore selection rule (\ref{paul1}) introduced by Pauli as the physical requirement is not a physical condition that must be necessarily satisfied.

Question is, if the mathematics used in verifying the vanishing of commutator as considered in (\ref{2.2N}) is correct and well defined, than how to interpret observation that relations (\ref{defPs}) require that selection rules (\ref{paul1}) are necessary to ensure the vanishing of  $[\hat M^2,\,M_x]$. To answer this let us recall that Pauli considers the case of a half integer value $m=1/2$ \cite{Pauli}. %When $m=1/2$, for the eigenfunction of $\hat M^2$ and $M_z$ one can choose  regular or singular function. \footnote{\bf{\color{red} WHY NECESSARILY REGULAR OR SINGULAR?}}  
For $m=1/2$, in calculating matrix element (\ref{defPs}) one term contains both regular and singular state vectors (one  term in $\hat M_x|m=1/2\rangle$ will be $|m=-1/2 \rangle$). In this case %is chosen to be regular and another to be singular, \footnote{\bf {\color{red} WHY SHOULD I CHOOSE ONE REGULAR AND ANOTHER SINGULAR? BECAUSE OPERATOR ALWAYS TURNES REGULAR INTO SUNGULAR?}} 
the so-called surface term arisies in integration. This term appears because the operator $\hat M^2$ in (\ref{defPs}) acts on both bra and ket vectors, in other words the surface term describes the difference between the  $\hat M^2 |\rangle$ and $\langle| \hat M^2$. These surface terms do not vanish and they guarantee that the matrix elements of the fundamental commutation relations $\left[\hat M^2,\hat M_k\right]$ are fulfilled without demanding the validity of the selection rules (\ref{paul1}).

We demonstrate this feature on the simpler example on the matrix elements of $\left[\hat M_z,\hat\phi\right]=-i$. As we have showed in a previous section, when th eigenfunctions are chosen as $\Psi(\phi|m)=e^{im\phi}$ operator $\hat M_z=-i\partial/\partial \phi$ is Hermitian provided that the spectrum satisfies condition $m-m'$ = integer. $\hat \phi$ is also Hermitian:
\begin{equation}
\label{phi}
\int^{2\pi}_0 \,d\phi\left[\Psi(\phi|m)\right]^{\star}\,\phi\,\Psi(\phi|m)= \int^{2\pi}_0 \,d\phi\left[\phi\,\Psi(\phi|m)\right]^{\star}\,\Psi(\phi|m).
\end{equation}
Though both operators are Hermitian and there are no singularities in space spanned by the eigenfunctions $\langle m|\phi \rangle =\Psi(\phi|m)$, evaluating matrix element as
\begin{equation}
\label{koba1}
\langle m|\hat M_z\hat\phi-\hat\phi\hat M_z|m' \rangle =m\langle m|\hat \phi |m' \rangle - m'\langle m|\hat \phi |m'\rangle=(m-m')\langle m |\hat \phi|m' \rangle,
\end{equation}
since this contradicts to $\left[\hat M_z,\hat \phi\right]=-i$. Indeed, straightforward calculation shows:
\begin{eqnarray}
(m-m')\langle m |\hat \phi|m' \rangle &=& (m-m')\int^{2\pi}_0 d\phi e^{-im\phi}\phi e^{im'\phi}=\nonumber\\
&& 2\pi i - (m-m')^{-1}\left[ e^{i 2\pi (m'-m)}-1\right] \neq -i\langle m|m'|\rangle.
\label{koba2}
\end{eqnarray}
It is easy to show that when  $m-m'$ is integer, $m-m'=k$,
\begin{equation}
\label{koba3}
(m-m')\langle m |\hat \phi|m' \rangle|_{m-m'=k} = 2\pi i \left(1-\delta_{0(m-m')}\right)=2\pi i - i\langle m|m'\rangle\neq -i\langle m |m'\rangle.
\end{equation}
The error can be fixed realizing that even in the case $m-m'$ integer the relation $\langle m|\phi \hat M_z|m'\rangle=m'\langle m|\phi|m'\rangle$ is  correct, and the relation $\langle m|\hat M_z\phi|m'\rangle= m\langle m|\phi|m'\rangle$ is not. Indeed:
\begin{eqnarray}
\langle m|\hat M_z\phi|m'\rangle &=& \int^{2\pi}d\phi \left[e^{im\phi}\right]^{\star}(-i\partial/\partial \phi)\phi e^{i m'\phi}=-2\pi i e^{2\pi i (m-m')}+\nonumber\\
&&\int^{2\pi}_0 d\phi \left[(-i\partial/\partial \phi)e^{im\phi}\right]^{\star}\phi. e^{im'\phi} = -2\pi i + m\langle m|\phi|m'\rangle.
\label{koba33}
\end{eqnarray}
Using (\ref{koba3}) and  (\ref{koba33}) we obtain equation which, in distinct of Eq.~(\ref{koba1}), respects fundamental commutation relation $\left[\hat M_z, \hat \phi\right] = -i$:%1footnote{{\bf{\color{red} How the right hand side is $-i\delta_{mm'}$? Can't derive.\\Is it $-i\delta_{mm'}$ or $-2\pi i\delta_{mm'}$? When $m=m'$, I am getting $-2\pi i$, not $-i$.}}}
\begin{equation}
\label{koba4}
\langle m|\hat M_z\hat\phi-\hat\phi\hat M_z|m' \rangle=-2\pi i + m \langle m|\phi|m'\rangle - m'\langle m|\phi|m'\rangle=-i\delta_{0,(m-m')}.
\end{equation}
An example above demonstrates that  the fundamental commutation relations in Hilbert space and for matrix elements can be and are satisfied without invoking selection rules like ones postulated in \cite{Pauli}.
Pauli used selection rule (\ref{paul1}) to show that the eigenvalues of angular momentum could be only integer and half-integer \cite{Pauli} and indeed if start from (\ref{paul1}) the eigenfunctions with the non integer eigenvalues do not appear. As we have shown in \cite{JKT}, non-integer eigenvalues are admissible from the theoretical viewpoint. Since (\ref{paul1}) is not a physical condition, dropping it does not lead to any theoretical contradiction or inconsistency.

Let us discuss the measurability of the components of the 
%What can be said regarding their physical character about the components of the 
angular momentum operator. When the complete set of commuting operators is given by $\hat M^2$ and $\hat M_z$ this means that in the corresponding scheme of measurement their corresponding eigenvalues are exactly measurable. 
In the same scheme $\hat M_x$ and $\hat M_y$ operators are not measurable quantities and therefore imposing with their help any conditions on eigenfunctions of $\hat M^2$ and $\hat M_z$
is as inconsistent as imposing the condition of single valuedness on the wave function. On the other hand, in the framework of quantum mechanics it is possible to prepare 
such a scheme of measurements in which, within some uncertainty, eigenvalues of non-commuting operators are also measurable. 
Such states in quantum mechanics are called mixed states and they have the same physical status as the pure states which are built using eigenstates of commuting operators \cite{shiff}-\cite{Landau}. According to quantum mechanics the mixed states are 
represented by certain superpositions of normalizable pure wave functions. 
In this case the components of the angular momentum are required to satisfy the following condition is imposed on all measurable quantities:
\begin{equation}
\langle \Psi^R_{Mm}(L';m') |\hat M_k|\Psi^R_{Mm}(L;m) \rangle < \infty;
\label{eq2.6n}
\end{equation}
This condition is satisfied not only for regular functions corresponding to integer-valued spectrum but also for regular functions corresponding to non-integer-valued spectrum. It is not difficult to demonstrate this statement explicitly. Indeed the regular eigenfunctions of the square of the angular momentum and its third component are orthogonal for integer-valued as well as non-integer-valued spectrum: 
\begin{equation}
\langle \Psi_{Mm}(L;m) |\Psi_{Mm}(L';m') \rangle =\int_{-1}^1 d \xi \int_{0}^{2\pi} d \phi [\Psi^R_{Mm}(\phi;\xi | L;m)]^* \Psi^R_{Mm}(\phi;\xi | L';m') =\delta_{L L'}\delta_{m m'};
\label{eq2.7n}
\end{equation}
Correspondingly, when sandwiched between the mixed states, $\hat M_x$ and $\hat M_y$ will have non-vanishing matrix elements only in case when $m'=(m\pm 1)$.
Indeed, in this case by acting with $\hat M_x$ and $\hat M_y$ on 
$\Psi^R_{Mm}(\phi;\xi |L';m')$ two terms are generated one of which is $\Psi^R_{Mm}(\phi;\xi |L';m)$, i.e. one term is the eigenfunction with the same $m$. For $m'\neq m\pm 1$ such a term does not appear. Therefore in case when $m'$ does not satisfy the above mentioned selection rule the matrix element $\langle \Psi_{Mm}(L';m') |\hat M_k|\Psi_{Mm}(L;m) \rangle$ vanishes due to orthogonality (\ref{eq2.1}).%will disappear due to the orthogonality connected to integration over $\phi$. 
When $m'=m\pm 1$,  term  $\Psi^R_{Mm}(\phi;\xi |L';m)$ generated from $\hat M_x \Psi^R_{Mm}(\phi;\xi |L';m\pm 1)$, in case of integer eigenvalues is regular, therefore the condition of Eq.~(\ref{eq2.6n}) is trivially satisfied. 

For non-integer eigenvalues the function $\Psi_{Mm}(\phi;\xi |L;m)$ may be regular as well as singular. 
For regular functions the condition of Eq.~(\ref{eq2.6n}) is again trivially satisfied, analogously to the case of integer eigenvalues. For the singular case, when one out of two terms of $\hat M_k|\Psi^R_{Mm}(L;m) \rangle$ is a singular function,  let us recall 
that the degree of singularity of $\Psi_{Mm}(\phi;\xi|L;m)$ is given by factor $(1-\xi^2)^{-|m|/2}$ \cite{JKT}. In the matrix element (\ref{eq2.6n}) this singulariryis fully regularized by the factor $(1-\xi^2)^{|m|/2}$, present in $[\Psi^R_{Mm}(\phi;\xi |L;m)]^*$. Therefore the condition of 
Eq.~(\ref{eq2.6n}) is satisfied for integer as well as for non-integer eigenvalues. Evidently, condition analogous to Eq.~(\ref{eq2.6n}) is guaranteed for any integer powers of $\hat M_x$ and $\hat M_y$ and for their linear combinations.

Thus we are led to the conclusion that the condition, "...by applying components 
%by acting with the components 
of the angular momentum $\hat M_k$ ($k=x,y,z$) to 
a given system of regular (i.e. quadratically integrable) eigenfunctions of the square of angular operator $\hat M^2$ corresponding to a fixed eigenvalue we should not leave this system of eigenfunctions ... Of course it is implicitly implied in our condition that the functions obtained by acting 
with $\hat M_k$ must be also regular." \cite{Pauli}, does  not single out the integer eigenvalues of the angular momentum as preferable ones.

\section{Conclusions}
\label{secVI}
%Let us summarise the obtained results:

%1begin{enumerate}
%\item

\noindent The eigenvalue equation for $\hat M^2$ is the second order linear homogenous differential equation and therefore it has two linearly independent solutions. Any 
%Because of linearity any 
linear combination of two concrete solutions is also a solution. In \cite{JKT} we showed that besides the solution corresponding to the only integer spectrum of eigenvalues, there also exists regular, i.e. physically admissible solution with the non integer eigenvalues of $\hat M^2$.%As a result of our investigation it became clear that among these linear combinations there are some for which the physical requirement of normalizability leads to integer-number spectrum of eigenvalues, however there are also other solutions which can be normalised for a non-integer eigenvalues.
\footnote{Unfortunately, during the creation of quantum mechanics solutions in the form of Legendre polynomials which lead to only integer values of the spectrum have not been critically analysed. As a result the possibility of the existence of solutions with the non integer eigenvalues has been completely excluded.}\\

\noindent Using commutation relations of the angular momentum operators and %certain 
an auxiliary postulate, not following and independent from the commutation relations, it has been shown that the only integer-valued spectrum of the square of the angular momentum and its third component 
is compatible with the algebra of commutators (see, e.g. \cite{Weinberg2015}). 
This postulate states that by acting repeatedly with the rising (lowering) operator on a  normalizable eigenfunction of  $\hat M^2$ and $\hat M_z$ with the negative (positive) eigenvalue $-m$ ($+m$), %negative (positive) eigenvalues of $\hat M_z$, 
one always  arrives to a normalizable eigenfunction with the positive (negative) eigenvalue $+m$ ($-m$). %corresponding to positive (negative) eigenvalues.
By explicitly solving eigenvalue equations we have shown that in general this is not the case  and therefore, this postulate does not constitute to a physical requirement \cite{JKT}.
 Thus the integer valued as well as non-integer valued spectrum of the eigenvalues of the angular momentum is compatible with the commutation relations.\\

\noindent Integer valued spectrum is also obtained demanding single valuedness of the eigenfunctions of $\hat M_z$. 
According to the widespread view, condition of single valuedness is related to the invariance of the observable physical quantities under rotations on $2\pi$ (see, e.g. \cite{shiff}-\cite{Landau}). The non single valuedness of the eigenfunction of $\hat M_z$ is usually discussed when the eigenfunction is written in spherical coordinates. 
We have shown that, from purely mathematical viewpoint, this eigenfunctions are in fact single valued functions.
The eigenfunction expressed in Cartesian coordinates indeed is not non single valued and this is due to the non single valuedness of power functions of complex variables. However, this feature is fully compatible with the well known fact that the wave function is defined up to an arbitrary phase.\\

\noindent The requirement of single valuedness and periodicity of wave function was rejected  by Pauli \cite{Pauli}-\cite{merz} as requirement imposed on non observables; we completely agree with his critique. Pauli introduced an alternate requirement from which he derived that the eigenvalues could be only integer \cite{Pauli}. We found out that this alternate requirement %By this argument we supported the analogous critical point of view by W. Pauli. Simultaneously, from the point of view of principles of quantum mechanics we criticised the new alternative requirement  introduced by Pauli that by acting with operators $\hat M^\pm$on normalisable eigenfunctions one should not obtain functions from the class of non normalisable functions. Based on the explicit solutions of the eigenvalue equations we have shown that this requirement by Pauli
 does not correspond to main physical principles of quantum mechanics 
similarly to the requirements of single valuedness and periodicity, which were rightfully criticised and rejected by Pauli himself.\\

\noindent Conclusion based on the results obtained in \cite{JKT} and in the present article is as follows:\\[0.1in]
\noindent{\it There exist regular, physically admissible solutions to eigenvalue equation corresponding to a non integer spectrum of eigenvalues of angular momentum. There is no any requirement in the framework of quantum mechanics that excludes solutions with the non integer spectrum, therefore, the issue of the nature of eigenvalues forming the spectrum of the angular momentum remains open.\\
\noindent What we can claim for sure within the range of physically acceptable arguments is  that to a fixed value of the square of the angular momentum indeed corresponds a discrete spectrum of eigenvalues of the third component of the angular momentum, defined by the relation $|m|=L-k,k=\{0,1,\cdots , [L] \}$. This relation is the result of a physical requirement of normalizability of a wave function. As for the eigenvalues of $L$, they can be integer as well as non-integer.}\\

\noindent Relation $|m|=L-k$, $k=0,1,...,[L]$, following from the requirement of normalizability, manifests the symmetry $\hat M^2(m)=\hat M^2(-m)$. The positive $(+m)$ and negative $(-m)$ eigenvalues are simultaneously generated from this relation and there is no need, as it is done in the analysis of algebra of commutation relations, to act repeatedly by rising (lowering) operator on $\Psi_{Mm}(L,-m)$ $(\Psi_{Mm}(L,m))$ in order to arrive to eigenfunction with negative (positive) eigenvalue \cite{Weinberg2015}. As demonstrated in \cite{JKT}, it is possible only for the integer $L$.  In general, for any $L$, the symmetry $\hat M^2(m)=\hat M^2(-m)$ guarantees that for a regular $\Psi_{Mm}(L,m)$ there always exist regular counterpart $\Psi_{M,m}(L,-m)$. Regular $\Psi_{Mm}$ contains absolute value of the parameter $m$, i.e.it is written as $\Psi_{Mm}(L,|m|)$. 
Condition of truncating hypergeometric series, i.e. reducing hypergeometric functions into polynomials that isolates regular eigenfunctions, guarantees that if $\Psi_{Mm}(L,|m|)$ is regular, so will be $\Psi_{Mm}(L,|m|-1)$. In other words, acting with the lowering operator $\hat M^-$ on $\Psi_{M,m}(L,|m|)$ results in both $\Psi_{Mm}(L,m-1)$ and $\Psi_{Mm}(L,-m-1)$. The latter are obtained acting with the lowering operator only once while in the framework of algebra of commutation relations it is postulated that acting repeatedly by lowering operator leads from $\Psi_{Mm}(L,m)$ to $\Psi_{Mm}(L,-m)$. This postulate is of purely mathematical origin, dos not follow from first principles and thus it is not necessary to hold. %it is not necessary in analysis of spectrum of observables. 
In \cite{JKT} we demonstrated that there are regular, physically admissible solutions of eigenvalue equation with non integer eigenvalue that violate the postulate of covering whole spectrum by moving up (down) with $\hat M^+$ $(\hat M^-$).\\

\noindent Discussion above calls to a question: how to explain or describe discreteness of atomic energy levels?
%Based on all that was said above a simple question raises: if the picture described by us indeed corresponds to physical reality then what happens to the problem of discrete atomic levels which are empirically observable? Such a question is indeed well grounded. 
In particular, the observed  discrete spectrum of hydrogen atom can be obtained in theoretical calculations if we solve the corresponding Schr\" odinger equation and assume/prove that the square of the angular momentum has a discrete spectrum of eigenvalues. Indeed, for $\varepsilon_k$,  the observable discrete energy levels of hydrogen atom, quantum mechanics yields $\varepsilon\sim (k+L+1)^{-2}$, where from the requirement that eigenfunction must be regular it follows that  $k$ must be an integer number \cite{shiff}-\cite{Landau}. %For $L$ integer quantum mechanics reproduces the observed discrete spectrum, but if $L$ is non integer, the theory fails to describe this concrete observation.\\
\\
\noindent From our analysis it follows that $L$ and $m$ differ by integer number, $|m|=L-k$, however, $L$ can take any value. In a theoretical scheme which would guarantee the discrete values of $L$ %it would be necessary that for eigenvalues of the square of the angular momentum 
there should exist relation of discreteness similar to the one which is obtained when indicating its relation to $m$. As such a relation does not exist for $L$, the only condition leading to a discrete $L$ was the statement that $L$ must be integer. %Presumably that is the reason why such arguments have been found which, although not fully compatible with physical principles, 
This would guarantee that theory describes well the discrete energy levels. %Indeed, for $\varepsilon_k$,  the observable discrete energy levels of hydrogen theory, quantum mechanics yields $\varepsilon_k\sim (n+L+1)^{-2}$, where from the requirement that eigenfunction must be regular it follows that  $k$ must be an integer number \cite{shiff}-\cite{Landau}. 
For $L$ integer quantum mechanics reproduces the observed discrete spectrum, but if $L$ is non integer, the theory fails to describe this concrete observation.\\
\noindent To avoid this discrepancy and at the same time to not to introduce the statement - $L$ is always integer - which does not follow from the first principles, in this concrete problem of describing the discrete energy levels we can simply swap the cause and the effect. Namely, %can claim for sure within the range of physically acceptable arguments that to a fixed value of the square of the angular momentum indeed corresponds a discrete spectrum of eigenvalues of the third component of the angular momentum, defined by the relation $|m|=L-k,k=\{0,1,\cdots , [L] \}$. As for the eigenvalues of the square of the angular momentum, they can be integer as well as non-integer. theoretical justification of the empirical discrete atomic levels. 
%This not quite consequential physical picture can be easily corrected. For this we will need to exchange the 
%cause and the result. 
%while  till now the discrete energetic spectrum was obtained from the discrete integer-valued spectrum of the square of the angular momentum, we can do the other way round: 
in the theoretical description %modeling of quantum mechanics 
the discrete spectrum of the angular momentum can be related the empirically observed discrete spectrum of atomic energy levels: since quantum mechanics leads to $\varepsilon\sim(k+L+1)^{-2}$, in order to satisfy the observation that $\varepsilon$ is discrete, one has to conclude that in this concrete problem $L$ has to be integer. In this case the quantum mechanical problem of the angular momentum will be analyzed using exactly the same mathematical instruments as before, however the discrete character of $L$ will be related to the empirical facts and not to the theoretical arguments. \\
\noindent From the physical point of view such a scheme will be more systematic and self-consistent. 
\\[0.1in]
\noindent We are indebted with J.~T.~Gegelia for useful discussions and critically reading the manuscript.

%\end{references}

\end{document}